\documentclass[prc,aps,showpacs,nofootinbib,twocolumn]{revtex4}
\pdfoutput=1 
\usepackage{amsmath}
\usepackage{amsfonts}
\usepackage{graphicx}
\usepackage{dcolumn}
\usepackage{hyperref}

\begin{document}

\include{defs}

\title{Multipole analysis of kaon photoproduction data}
\author{Ron L.\ Workman}
\affiliation{ Center for Nuclear Studies,
Department of Physics, The George Washington University,
Washington, D.C. 20052}

\date{\today}

\begin{abstract}
Methods used in the multipole analysis of low-energy
pion photoproduction data have been tested against the database
of kaon photoproduction measurements. Results for some
multipoles are in
qualitative agreement with existing phenomenological models,
while others are unstable, given the present database.
These findings are compared to those of previous studies.

\end{abstract}


\maketitle

\section{Introduction}
\label{sec:intro}
The study of kaon photoproduction is motivated by the possibility of
a model-independent amplitude reconstruction~\cite{complete}. This
type of analysis has been realized for nucleon-nucleon 
scattering~\cite{nn} and may soon be attempted for kaon photoproduction,
given the existing
and expected data for single- and double-polarization quantities.
A second motivation hinges on the possibility that a resonance analysis
will reveal states not clearly seen in processes containing initial
or final single-pion states. Several recent~\cite{lotharme,grme} and
older~\cite{grushin,omelaenko} studies have noted that these two
goals (amplitude reconstruction and multipole analysis) are very 
different in nature. In the following, we have applied the method
used in Ref.~\cite{grme} to the low-energy kaon photoproduction
database in order to do a multipole analysis. 

In Ref.~\cite{grme}, the low-energy pion photoproduction database
was analyzed using a two-step procedure. Charged-pion photoproduction
data were fitted assuming a known t-channel pion exchange mechanism
for the high (unfitted) multipoles. This effectively removed the
overall phase ambiguity that arises when only a finite number of 
multipoles are fitted, and the higher waves are neglected. 
The obtained charged-pion multipoles, and
an appliction of Watson's theorem~\cite{watson} for multipoles 
coupled to the elastic $P_{33}$ $\pi N$ partial wave, allowed for the
determination of neutral-pion production multipoles. In a similar
approach, not using Watson's theorem directly, 
Grushin~\cite{grushin} was able to extract the 
$\pi N$ phases from a fit to photoproduction data.

As charged-kaon photoproduction has a t-channel process
contributing to high partial waves, the method used in Ref.~\cite{grme}
was applied over a region where $s-$ and $p-$wave multipoles were
expected to dominate. We have focused on energies from threshold
to 1.8 GeV in center-of-mass energy ($W_{CM}$), as this covers a
region influenced by the N(1650)$S_{11}$, N(1710)$P_{11}$, and
N(1720)$P_{13}$ resonances, included in the kaon MAID fit, but is below
the proposed N(1900)$D_{13}$~\cite{kmaid,tmart}.   

The t-channel process in kaon photoproduction
is generally~\cite{kmaid,tmart} 
built up from the exchange of the $K^+$, $K^*(892)$, and
$K^*(1270)$. The latter two states have sizeable widths producing
complex multipoles. The contribution to the total cross section from
Born plus $K^*$ exchange diagrams is known to increase rapidly at moderate
energies and this feature is usually handled with the introduction of
form factors~\cite{ff}. 
As a result, the prediction of higher multipoles is more
model-dependent here than in the case of charged-pion photoproduction.
 
We should also note that a recent single-energy multipole analysis 
has found broad
bands of multipole solutions, which are experimentally indistinguishable,
using the existing database~\cite{sandorfi}. 
Our findings are slightly more optomistic, showing at least qualitative 
agreement with some previous phenomenological fits. Problems noted in
Ref.~\cite{sandorfi} have been found here as well. 

\section{Fitting the Data}

Data were fitted at single energies corresponding to the
low-energy measurements
of $P$, $\Sigma$, $T$, $O_x$, and $O_z$ from 
the GRAAL collaboration~\cite{graal}. Differential cross sections
were taken from the CLAS collaboration~\cite{clas}. The lowest
energy point from Ref.~\cite{graal} was omitted, as contradictory
differential cross section measurements exist just above and below
this energy. Conventions for the signs of polarized measurements
were taken from Ref.~\cite{lotharme,bds}, differing from those
adopted in Ref.~\cite{sandorfi}.

In fitting the data, multipoles up to $E_{2-}$ and $M_{2-}$ were
varied. Waves beyond those being fitted were taken from the Born plus
$K^*$ exchange diagrams used in the kaon MAID fit~\cite{kmaid,tmart}.
In practice, however, over the considered energy range, the
$E_{2-}$ and $M_{2-}$ multipoles were found to be compatible
with the Born contribution, apart from contributions to the
real parts of $E_{2-}$ and $M_{2-}$. The resulting values for chi-squared
per degree of freedom were between 0.5 and unity for all searches.

In Fig.~1 we show the quality of our fit to the cross section, $P$,
$\Sigma$, and $T$ data, corresponding to lab photon energy of 1027 MeV
($W_{cm}$=1675 MeV).
Figure 2 shows the fit to $O_x$ data and a prediction of 
$O_z$, $C_x$, and $C_z$. Fitting $O_z$ and predicting the remaining
observables gives a comparable result. 
The results in Fig.~3, for beam-target quantities E,F,G and H, are
predictions. Preliminary CLAS data for E~\cite{pasyuk} show agreement
with this behavior, but rather poor agreement with other previous analyses.

The level of agreement, only
slightly worse than a fit of all observables, provides a 
validation of the method. However, the problem of uniqueness for
extracted multipoles, found in Ref.~\cite{sandorfi}, still persists. 

\begin{figure*}[h]
\includegraphics[width=0.60\textwidth, angle=90]{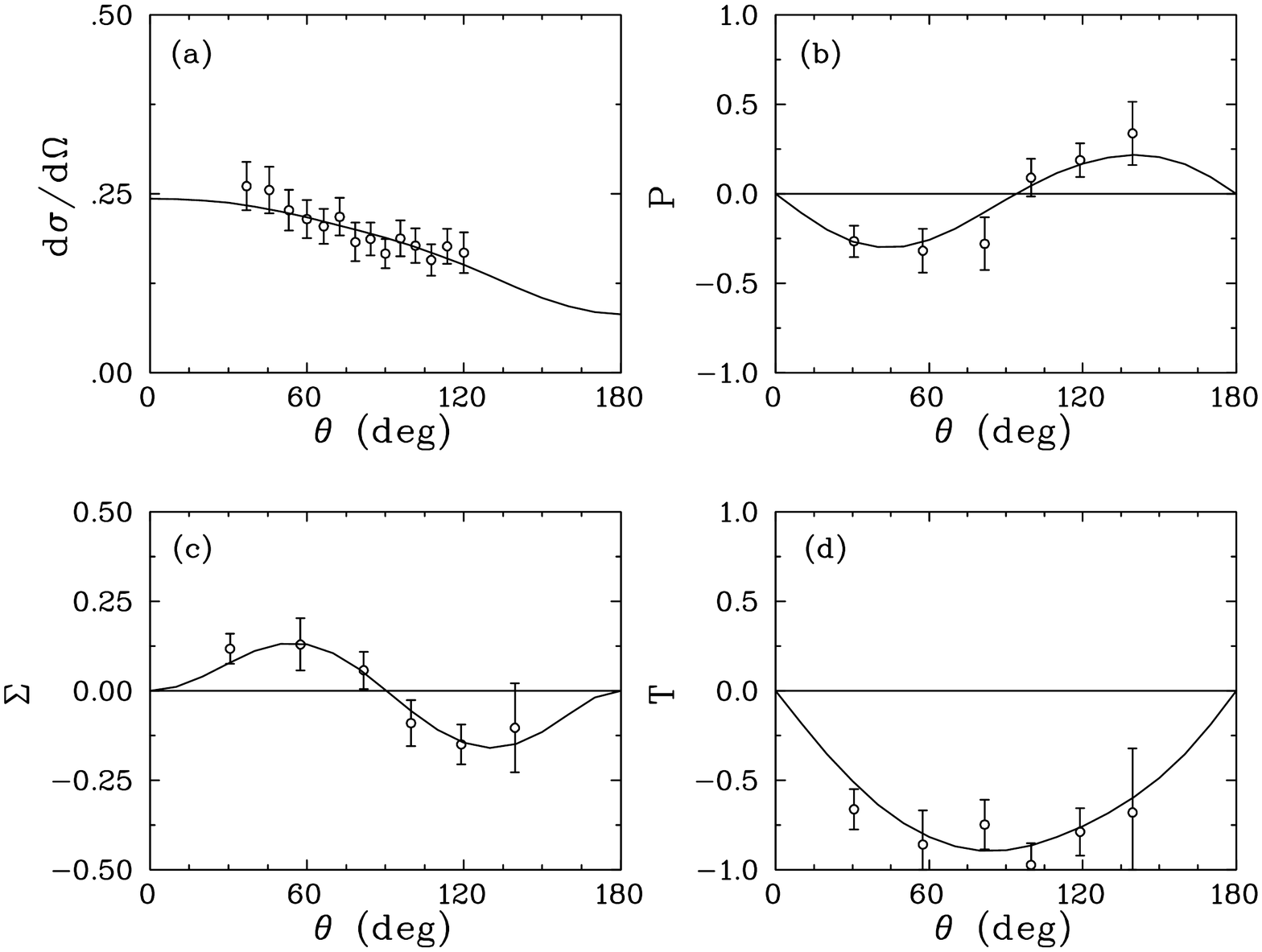}
\caption{\label{fig:dpst} Fit to $\gamma p\to K^+ \Lambda$ 
data: $d\sigma /d\Omega$ (a), $P$ (b), $\Sigma$ (c), and $T$ (d) 
for $E_{\gamma}$=1027 MeV. Data from Refs.~\cite{graal,clas}.}
\end{figure*}

\begin{figure*}[h]
\includegraphics[ width=300pt, angle=90]{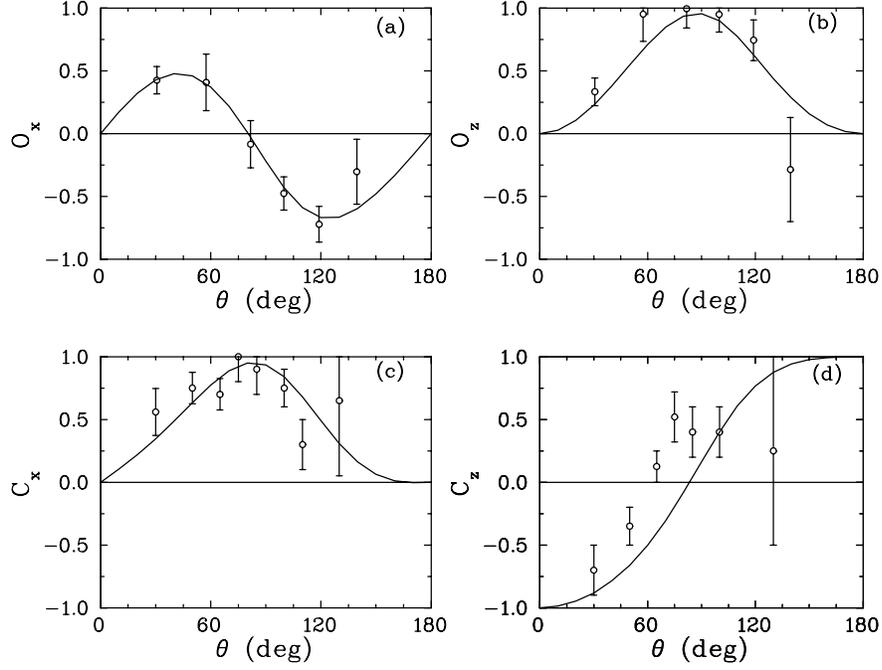}
\caption{\label{fig:oxoz} 
Fit to $\gamma p\to K^+ \Lambda$ data $O_x$ (a), and predictions for
$O_z$ (b), $C_x$ (c), and $C_z$ (d) 
for $E_{\gamma}$=1027 MeV. Data from Refs.~\cite{graal,sandorfi}.
} 
\end{figure*}

\begin{figure*}[h]
\includegraphics[ width=300pt, angle=90]{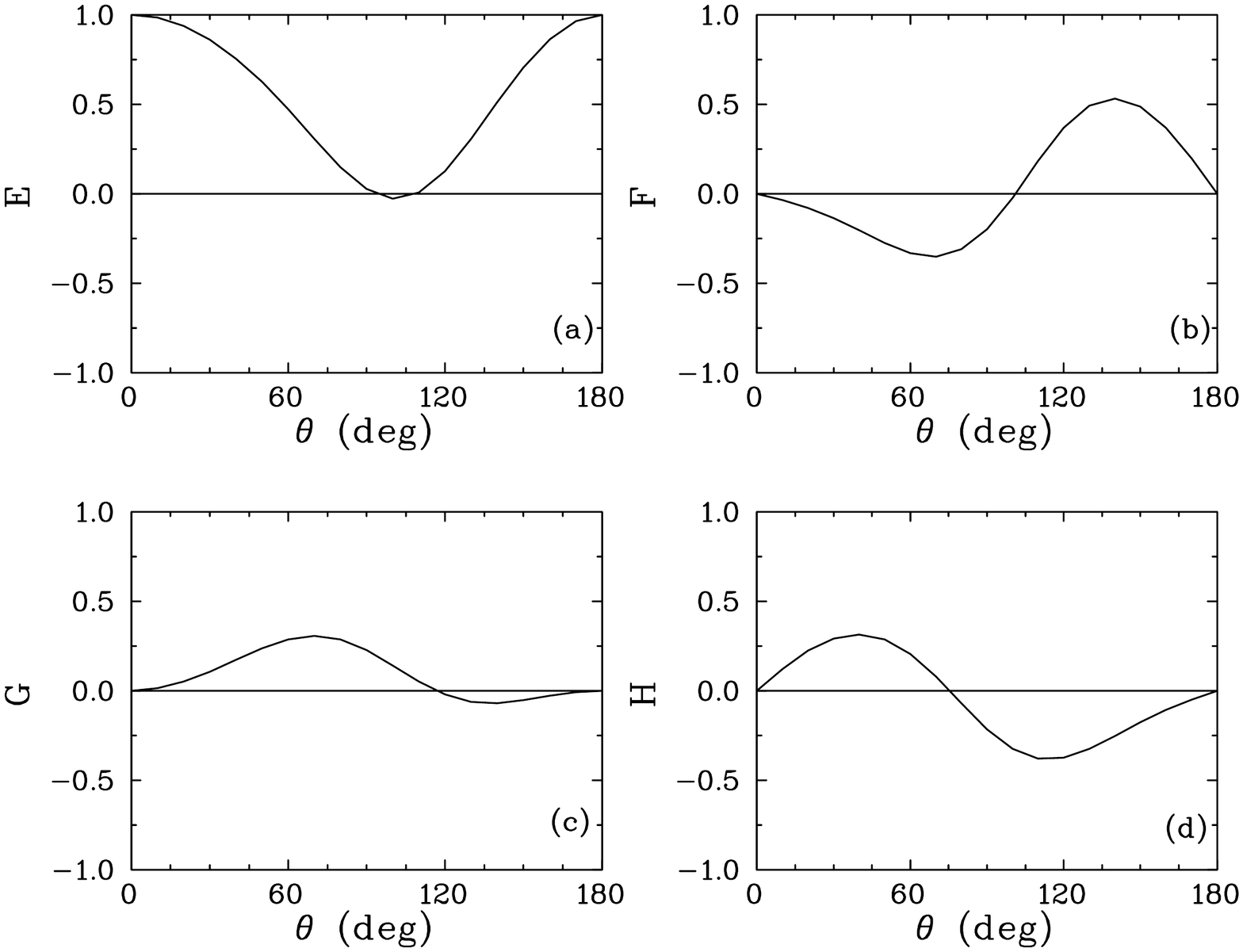}
\caption{\label{fig:efgh}
Predicted $\gamma p\to K^+ \Lambda$ data E (a), F (b), G (c), and H (d) 
for $E_{\gamma}$=1027 MeV. 
}
\end{figure*}

\section{Multipoles}

In Fig.~3, a set of single-energy multipoles is compared to the 
kaon MAID solution. The agreement is surprisingly good,
particularly for the imaginary parts, which appear qualitatively
consistent with the N(1650) and N(1710) resonances contained
in kaon MAID. The behavior of $E_{1+}$ is also evidence for
a contribution from the N(1720). However,
the $M_{1+}$ multipole has a 
large real part not included in kaon MAID. A feature quite similar
to this is seen in the energy-dependent multipole
analysis of Mart~\cite{tmart}. A large near-threshold value,
with a different sign, is also found in the Bonn-Gatchina
analysis~\cite{BoGa}.

In Fig.~3, the multipoles at a particular energy 
were obtained through searches in parameter space
near adjacent-energy solutions. However, in expanding the parameter
space, we found indistinguishable solutions, with very low chi-squared, 
corresponding to different multipole sets. Results for the $E_{1+}$ and
$M_{1-}$ were not stable. However, the trend seen in the $E_{0+}$ and
$M_{1+}$ multipoles remains in these alternate solutions. 
Similar difficulties were also found in the multipole analysis 
of Ref.~\cite{sandorfi}. 

In Fig.~4, we show the dominant multipoles fitted up to 1.9 GeV. The
fit remains quite good, retaining a chi-squared per degree of freedom
less than unity, without requiring a search of higher multipoles. The 
real part of the $M_{1+}$
multipole shows a rapid drop to zero, associated with a
rise in the imaginary part, at approximately 1.9 GeV. This supports
the claim for a 2-star N(1900) contribution~\cite{BoGa}.

\section{Conclusions}

We have performed an exploratory single-energy multipole analysis of
kaon photoproduction data near threshold - a region
where we assumed only a few multipoles would deviate from Born predictions.
The results are mixed. We have been able to fit a
minimal set of data, predicting other observables, thus testing the
method. However, a unique set of multipoles was not found. This confirms
the findings of Ref.~\cite{sandorfi} with a somewhat different method.

Clearly, the validity of this method rests on the model-independence
of the t-channel process. It would be useful to repeat these fits
with a set of different models for this contribution. Many recipies
exist to moderate the Born contributions. Their
effect is likely to add another source of model-dependence with increased
energy. The fact that our search tended to require corrections to
the real parts of the $E_{2-}$ and $M_{2-}$ multipoles suggests that
a modified t-channel contribution may be required.
This issue is presently being studied. Numerical values for
the single-energy solutions and included datasets will be provided
through the SAID webpage~\cite{SAID}.

Plots of the extracted multipoles show qualitative similarities to
the kaon MAID fit in some cases near threshold. For the $M_{1+}$,
however, the result is qualitatively different near threshold, with
resonance-like behavior near the N(1900) resonance energy. 

In Ref.~\cite{sandorfi}, 
multiple solutions were also found in fits to a larger set of 
observables, with a low chi-squared. 
Multipoles up to L=3 were searched, whereas we have taken waves with
L$>$1 to be given by the Born contribution,
including $K^*$ exchange, with corrections to the real parts of $E_{2-}$ and
$M_{2-}$ contributions found necessary in the fit. Fitting higher multipoles 
decreases the chi-squared but also reduces
the Born contributions, which serve to remove
the overall phase ambiguity. We have also found that including
the $E_{2-}$ and $M_{2-}$ multipoles in the fit, when not 
required, tended to increase the uncertainties on the other fit parameters.   

Multipole analyses come in two main types: those with a significant
t-channel component, which is taken as known to remove the
overall sign ambiguity (such as $\pi^+ n$ photoproduction),
and others with a negligible t-channel contribution (such
as $\pi^0 p$ photoproduction at low energies). In the latter
case, one is free to fix one multipole phase and determine
multipoles only up to this unknown phase (which is a function
of energy). In Ref.~\cite{sandorfi}, Born terms were added
and a phase for the $E_{0+}$ multipole was chosen. This
was not done in the present fit. Here the Born contribution has fixed
our overall phase and therefore the phase of $E_{0+}$ as well.

We conclude by emphasizing that these single-energy multipole analyses are 
limited both by data quality and our confidence in the 
high angular momentum component.

\begin{figure*}
\includegraphics[ width=300pt, keepaspectratio, angle=90]{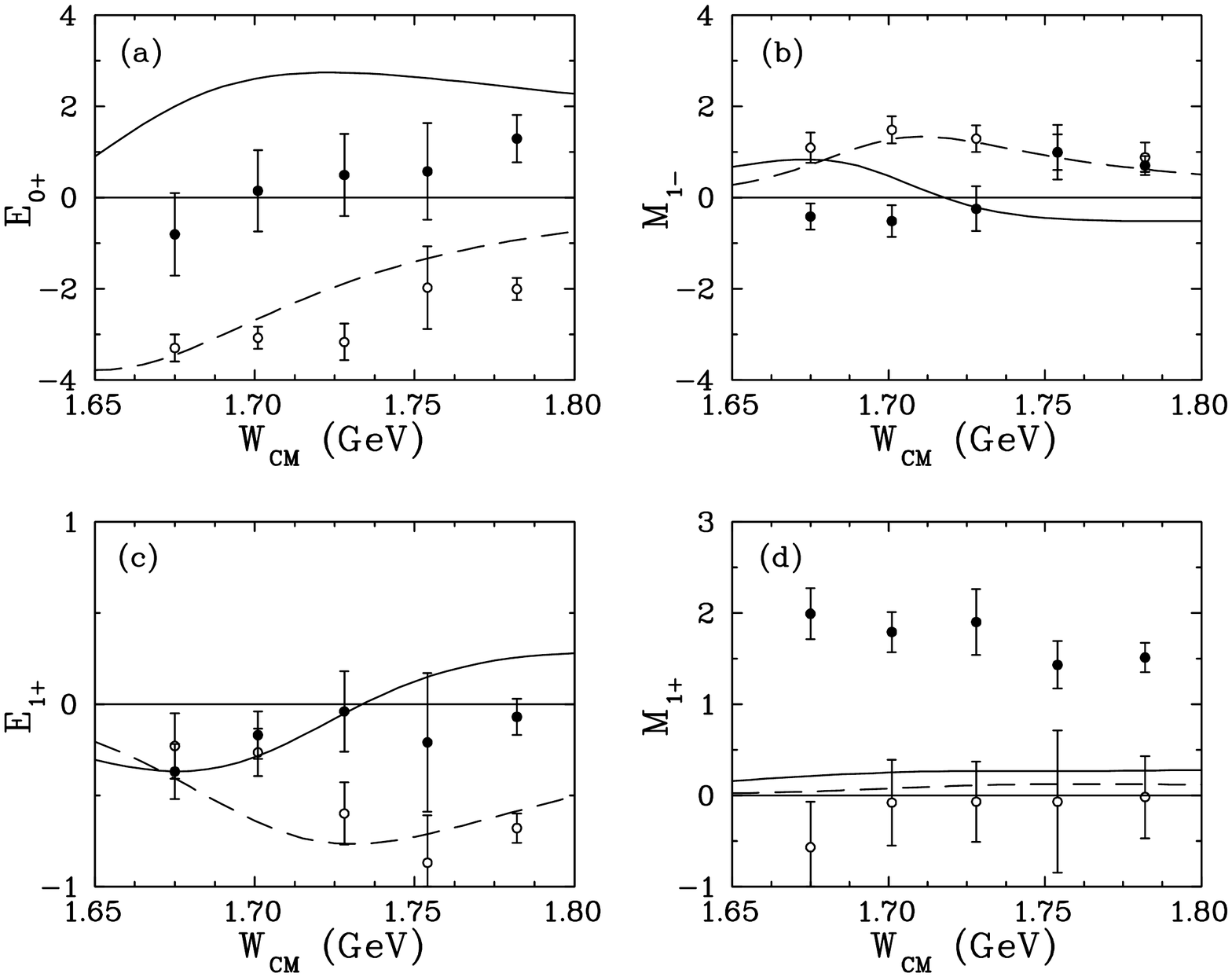}
\caption{\label{fig:mpoles} Single-energy multipoles $E_{0+}$ (a),
$M_{1-}$ (b), $E_{1+}$ (c), and $M_{1+}$ (d). Solid points
correspond to real parts and open points give the imaginary
parts of single-energy fits. Curves, solid (real) and dashed (imaginary),
are given by the Kaon MAID solution. Units are $10^{-3}/m_{\pi^+}$.
}
\end{figure*}

\begin{figure*}
\includegraphics[ width=300pt, keepaspectratio, angle=90]{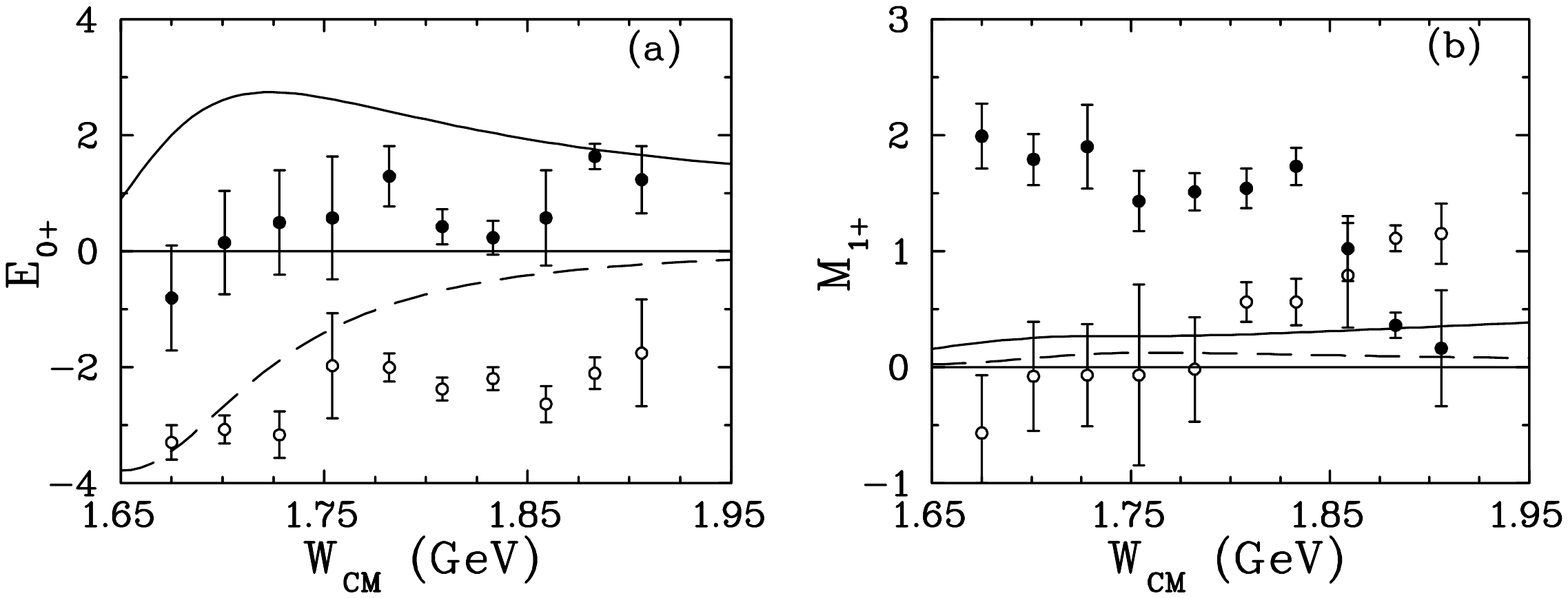}
\caption{\label{fig:fig2} Single-energy multipoles, $E_{0+}$ (a) and
$M_{1+}$ (b), extended to 1.9 GeV. Notation as in Fig.~4.
Units are $10^{-3}/m_{\pi^+}$.
}
\end{figure*}

\begin{acknowledgments}
The author thanks I.I. Strakovsky for updating the SAID database
for kaon photoproduction.
This work was supported in part by the U.S.\ Department of Energy
Grant DE-FG02-99ER41110. 
\end{acknowledgments}

\end{document}